\documentstyle[12pt,aps,psfig]{revtex}
\begin{document}
\tightenlines
\title{Transform of Riccati equation of constant coefficients through 
fractional procedure}
\author{H.C. Rosu$^1$\footnote{hcr@ipicyt.edu.mx}, A.L. Madue\~no$^2$, 
J. Socorro$^2$} 
\address{$^1$ Dept. of Appl. Math., IPICyT, Apdo Postal 3-74 Tangamanga, 
San Luis Potos\'{\i}, MEXICO\\ 
$^2$ Instituto de F\'{\i}sica, Universidad de Guanajuato, Apdo Postal E-143, 
Le\'on, MEXICO}

\maketitle
\begin{abstract} 
 We use a particular fractional generalization of the 
ordinary differential equations that we apply to the 
Riccati equation of constant coefficients. By this means the latter is 
transformed into a modified Riccati equation with the
free term expressed as a power of the independent variable which is of the 
same order as the order of
the applied fractional derivative. We provide
the solutions of the modified equation and employ the results for the 
case of the cosmological Riccati equation of FRW barotropic cosmologies 
that has been recently introduced
by Faraoni.
\end{abstract} 

\noindent
\centerline {{\bf J. Phys. A 36 (4), 1087-1094  (31 January 2003)}} 

PACS Numbers: 02.30.Hq; 04.20.Jb \hfill  arXiv Number: math-ph/0112020\\

\section{ Introduction}

The fractional calculus is a generalization of the ordinary
differential and integral calculus \cite{Ross}. The main point is how to 
think about
the derivative of order $r$, where $r$ is an arbitrary real or complex number.
In 1695, 
${\rm L'H\hat ospital}$ was the first to ask in a letter to Leibnitz on the
possibility to perform calculations by means of a fractional derivative of 
order $r=\frac{1}{2}$.
Leibnitz answered that the question looked as a  paradox to him but he 
predicted that in the future useful consequences might occur.
 In 1697, Leibnitz referring to the infinite product of 
 Wallis for $\pi / 2$ 
used the notation $d^{1/2}y$ and surmised that the fractional
calculus could be used to get the same result.

In 1819 the first mention of derivatives of arbitrary order
occurred in a published text. 
The French mathematician S.F. Lacroix published 700
pages on the differential calculus, where one can find 
less than 2 pages dedicated to the fractional topic, which seems to be
based on a result of Euler dated 1730.
He started with 
\begin{equation}
y=x^n~,
\end{equation}
where $n$ is an integer and  wrote the $m$th derivative in the form
\begin{equation}
\frac{d^{m}y}{dx^{m}}=\frac{n!}{(n-m)!}x^{n-m}~.
\end{equation}
Next, Lacroix changed the factorial using the $\Gamma (x)$ function 
(introduced by Legendre).  $n$ was changed from an integer to a real number, 
$n\rightarrow a$, 
and $m$ was chosen to be $m=\frac{1}{2}$; thus
\begin{equation}
\frac{d^{1/2}y}{dx^{1/2}}=\frac{\Gamma(a+1)}{\Gamma(a+\frac{1}{2})}
x^{a-\frac{1}{2}}~.
\end{equation}
In this way, he expressed the derivative of order one half by an arbitrary
power of $x$. A simple example given by Lacroix refers to the case $y=x$
\begin{equation}\label{L4}
\frac{d^{1/2}x}{dx^{1/2}}=\frac{2 \sqrt{x}}{\sqrt{\pi}}~.
\end{equation}
Along the years, great mathematicians, such as Euler, Fourier,
Abel and others, did some work on the fractional calculus that nevertheless
remained as a sort of curiosity. 

The modern epoch started in 1974, when a consistent formalism of
the fractional calculus has been developed by K.B. Oldham and 
J. Spanier \cite{OS}.
It has been found that Lacroix's result (\ref{L4}) coincides with that obtained by means of the 
present definition of the Riemann-Liouville fractional derivative. 

\section{ Basic definitions of  the fractional calculus}

\noindent
(i) One can define the {\it fractional integral} of order $\alpha >0 $ 
as follows

\begin{equation}
_{x_{0}}D_{x}^{-\alpha }f(x)=\frac{1}{\Gamma (\alpha )}\int_{x_{0}}^{x}
\frac{f(x^{\prime })dx^{\prime }}{(x-x^{\prime })^{1-\alpha }} . \label{c1}
\end{equation}

In particular, for $x_{0}=0$ one usually writes

\begin{equation}
D_{x}^{-\alpha }f(x)=\frac{1}{\Gamma (\alpha )}\int_{0}^{x}\frac{f(x^{\prime
})dx^{\prime }}{(x-x^{\prime })^{1-\alpha }}.
\end{equation}

\noindent
(ii) For $\beta \geq 0$ one can define the {\it fractional derivative}
of order $\beta$ in the following way

\begin{equation}
\frac{d^{\beta }f(x)}{dx^{\beta }}=D_{x}^{\beta }f(x)
=\frac{d^{n}}{dx^{n}}D_{x}^{-(n-\beta)}f(x)
=\frac{d^{n}}{dx^{n}}\frac{1}{\Gamma
(n-\beta )}\int_{0}^{x}\frac{f(x^{\prime })dx^{\prime }}{(x-x^{\prime
})^{1-n+\beta }}~,
\end{equation}
where $n\geq \beta$. 
Thus the $\beta$ fractional derivative is defined as an ordinary derivative of order $n$
of the fractional integral of order $n-\beta$.

\noindent
(iii) The {\em chain rule} has the form  
\begin{equation}
\frac{d^{\beta }}{dx^{\beta }}f(g(x))=\sum_{k=0}^{\infty }\left( ^{\beta }
_{k} \right)_{\Gamma} \left( \frac{d^{\beta -k}}{dx^{\beta -k}}1 
\right)\frac{d^{k}}{%
dx^{k}}f(g(x)),
\end{equation}
where $k\in N$ and $\left( ^{\beta } _{k} \right)_{\Gamma}$ are the 
coefficients of the generalized binomial  
\begin{equation}
\left( ^{\beta } _{k} \right)_{\Gamma}
=\frac{\Gamma (1+\beta )}{\Gamma (1+k)\Gamma
(1-k+\beta )}~.
\end{equation}
\bigskip

\noindent
(iv) {\em Leibnitz's rule} for the derivative of the product has the form
\begin{equation}
\frac{d^{\beta }}{dx^{\beta }}f(x)g(x)=\sum_{k=0}^{\infty }\left( ^{\beta }
_{k} \right)_{\Gamma}\frac{d^{k}}{dx^{k}}f(x)
\frac{d^{\beta -k}}{dx^{\beta -k}}g(x)~,
\end{equation}
where $k\in N$.

\section{ Ordinary differential equations from the 
fractional calculus}

\noindent
Let $\ \widehat{L}$ be a differential operator and let its action 
$\widehat{L}[u(x)]=g(x)$
be used to define a differential equation.
We generalize $\widehat{L}$ to the fractional calculus
by means of the fractional derivative 
of order $\beta =1-\delta$, $\delta \in (0,1)$, writing the following
 fractional differential equation
\begin{equation}
D_{x}^{1-\delta }\widehat{L}[u(x)]=g(x).  \label{e1}
\end{equation}
When $\delta =1$, then $D_{x}^{1-\delta }\widehat{L}[u(x)]=
\widehat{L}[u(x)]$.

One can obtain a solution of (\ref{e1}) by applying the fractional integral
of the same order to the left 
\begin{equation}
D_{x}^{-(1-\delta )}\left\{ D_{x}^{1-\delta }\widehat{L}[u(x)]\right\}
=D_{x}^{-(1-\delta )}[g(x)],  \label{e2}
\end{equation}
\begin{equation}
\widehat{L}[u(x)]=D_{x}^{-(1-\delta )}[g(x)],  \label{e3}
\end{equation}
\begin{equation}
\widehat{L}[u(x)]=D_{x}^{\delta -1}[g(x)].  \label{e4}
\end{equation}
For example: $\widehat{L}=\frac{d}{dx}+p(x)$ implies
\begin{equation}
\widehat{L}[u(x)]=\frac{d}{dx}u(x)+p(x)u(x),  \label{e5}
\end{equation}
\begin{equation}  \label{e6}
\frac{d}{dx}u(x)+p(x)u(x)=D_{x}^{\delta -1}[g(x)],
\end{equation}
then
\begin{equation}  \label{e7}
u(x)=\frac{1}{\mu (x)}\Bigg[ \int^{x}\mu
(s)\, D_{s}^{\delta-1}g(s)ds+c\Bigg],~
\end{equation}
where
\begin{equation}  \label{e8}
\mu (x)=\exp \left(\int^{x}p(s)ds\right)~.
\end{equation}
Equations (\ref{e7}) and (\ref{e8}) provide 
the solution to the linear generalized equation of the first order.
This is not the unique possible fractional 
generalization \cite{Eli,mgn}. One could have taken 
$D_{x}^{1-\delta }u(x)+p(x)u(x)=g(x)$ or some other procedure. However,
the present approach leads to analytical results in applications.

\noindent
\section{Application to the Riccati equation of constant coefficients}

The ordinary Riccati equation of constant coefficients is 
$\frac{du(x)}{dx}+au^2(x)=b$,
where $a$ and $b$ are constants. Thus the operator
of the Riccati type is 
$\widehat{L}_R=\frac{d}{dx}+au(x)$ acting always in the space of 
functions $u(x)$, i.e., $\widehat{L}_Ru(x)=\frac{du(x)}{dx}+au^2(x)=g(x)$.
Fractional considerations related to this operator can be found 
in the work of Metzler et al \cite{mgn}. 

The fractional Riccati equation according to the scheme proposed in the 
previous section is
\begin{equation}  \label{e9}
D_{x}^{1-\delta }\Bigg[ \frac{d}{dx}+au(x)\Bigg] u(x)=b,
\end{equation}
or
\begin{equation}  
\frac{du(x)}{dx}+au^2(x)=D_{x}^{-(1-\delta )}b. \label{e11}
\end{equation}
The right hand side can be written
\begin{equation}
b[D_{x}^{-(1-\delta )}1] 
=\frac{b}{\Gamma (1-\delta )}\int_{0}^{x}\frac{dt}{(x-t)^{\delta }} 
=\frac{b\, x^{1-\delta }}{\Gamma (1-\delta )(1-\delta )}  
=\frac{bx^{1-\delta }}{\Gamma (2-\delta )}~.
\end{equation}
Thus, solving the fractional Riccati equation is equivalent to solving
the following type of particular, ordinary Riccati equation that we call
the $\delta$-modified Riccati equation
\begin{equation}
\frac{du(x)}{dx}+a[u(x)]^{2}=\frac{bx^{1-\delta }}{\Gamma (2-\delta )}~.
\label{e14}
\end{equation}

\subsection{ Solution of the $\delta$-modified Riccati equation}

In order to solve $(\ref{e14})$ we use the transformation  
$u=y^{\prime }/ay$; $u^\prime=[y^{\prime \prime }y
-(y^{\prime })^{2}]/[ay^{2}]$. One gets
the associated linear second order differential equation
\begin{equation}
\frac{d^{2}y(x)}{dx^{2}}-\frac{ab}{\Gamma (2-\delta )}x^{1-\delta }y=0.
\label{e15}
\end{equation}
Multiplying by $x^2$ leads to
\begin{equation}
x^{2}\frac{d^{2}y(x)}{dx^{2}}-\frac{ab}{\Gamma (2-\delta )}x^{3-\delta }y=0.
\label{e16}
\end{equation}
The latter has solutions expressed in terms of Bessel functions. To see this 
we use the following known result. The equation
\begin{equation}
x^{2}y^{\prime \prime }+(1-2p)xy^{\prime }+\Bigg[q ^{2}r
^{2}x^{2r}+\left( p ^{2}-n^{2}r ^{2}\right) \Bigg] y=0
\label{e17}
\end{equation}
has 
for real $q$ the linear independent solutions
\begin{equation}
y_{1}=x^{p}J_{n}(q x^{r}),  \label{e19}
\end{equation}
\begin{equation}
y_{2}=x^{p}Y_{n}(q x^{r }),  \label{e20}
\end{equation}
where $\ J_{n}(x)$ and $Y_{n}(x)$ are the Bessel functions
of the first and second type, respectively. 
We shall also use the following properties of the Bessel functions
\begin{eqnarray}
J_{n}^{\prime }(x)&=&J_{n-1}(x)-\frac{n}{x}J_{n}(x)  \label{e21} \\
J_{n}^{\prime }(x) &=&\frac{1}{2}\Bigg[ J_{n-1}(x)-J_{n+1}(x)\Bigg]
\label{e22} \\
J_{n}^{\prime }(x)&=&\frac{n}{x}J_{n}(x)-J_{n+1}(x).  \label{e23}
\end{eqnarray}
The same holds for $Y_{n}(x)$. Then, using (\ref{e21}) one gets 
\begin{equation}
y_{1}^{\prime }=y_{1}\Bigg[\frac{p -nr }{x}+q r
x^{r -1}\frac{J_{n-1}(q x^{r})}{J_{n}(q x^{r})}%
\Bigg],  \label{e24}
\end{equation}
\begin{equation}
y_{2}^{\prime }=y_{2}\Bigg[\frac{p -nr}{x}+q r
x^{r -1}\frac{Y_{n-1}(q x^{r })}{Y_{n}(q x^{r })}%
\Bigg]~.  \label{e25}
\end{equation}
On the other hand, using (\ref{e23}) one gets
\begin{equation}
y_{1}^{\prime }=y_{1}\Bigg[ \frac{p +nr }{x}-q r
x^{r -1}\frac{J_{n+1}(q x^{r })}{J_{n}(q x^{r})}%
\Bigg],  \label{e26}
\end{equation}
\begin{equation}
y_{2}^{\prime }=y_{2}\Bigg[ \frac{p +nr }{x}-qr
x^{r -1}\frac{Y_{n+1}(q x^{r })}{Y_{n}(q x^{r })}%
\Bigg].  \label{e27}
\end{equation}
In order to find the Riccati solutions we can 
identify the parameters from comparison of (\ref{e16}) and (\ref{e17})
\begin{equation}
p =\frac{1}{2}~,\quad q =\frac{2}{3-\delta }\sqrt{-\frac{ab}{\Gamma (2-\delta )}}~,\quad r =\frac{3-\delta }{2}~,\quad
n =\frac{1}{3-\delta }~.  \label{e28}
\end{equation}
 It is more convenient to work with  equations (\ref{e24})  and 
(\ref{e25})
because $p -nr =0$. Therefore
\begin{equation}
u_{1}(x)=\frac{y_{1}^{\prime }}{ay_{1}}=\frac{1}{a}\Bigg[ q r x^{r -1}\frac{J_{n-1}(q
x^{r })}{J_{n}(q x^{r})}\Bigg]~,  \label{e29}
\end{equation}
\begin{equation}
u_{2}(x)=\frac{y_{2}^{\prime }}{ay_{2}}=\frac{1}{a}\Bigg[ q r x^{r -1}\frac{Y_{n-1}(q
x^{r })}{Y_{n}(q x^{r })}\Bigg]~,  \label{e30}
\end{equation}


and thus 
\begin{equation}
u_{1}(x)=\frac{1}{a}\Bigg[\sqrt{\frac{ab}{\Gamma (2-\delta )}%
}x^{\frac{1-\delta }{2}}\frac{J_{\frac{\delta -2}{3-\delta }}(\frac{2}{%
3-\delta }\sqrt{\frac{ab}{\Gamma (2-\delta )}}x^{\frac{%
3-\delta }{2}})}{J_{\frac{1}{3-\delta }}(\frac{2}{3-\delta }\sqrt{\frac{ab}{%
\left| \Gamma (2-\delta )\right| }}x^{\frac{3-\delta }{2}})}\Bigg]~,
\label{e35}
\end{equation}

\begin{equation}
u_{2}(x)=\frac{1}{a}\Bigg[ \sqrt{\frac{ab}{\Gamma (2-\delta )}%
}x^{\frac{1-\delta }{2}}\frac{Y_{\frac{\delta -2}{3-\delta }}(\frac{2}{%
3-\delta }\sqrt{\frac{ab}{\Gamma (2-\delta ) }}x^{\frac{%
3-\delta }{2}})}{Y_{\frac{1}{3-\delta }}(\frac{2}{3-\delta }\sqrt{\frac{ab}{%
\Gamma (2-\delta )}}x^{\frac{3-\delta }{2}})}\Bigg]~.
\label{e36}
\end{equation}
For positive $b$ one gets an imaginary $q$
parameter that turns the $J$ and $Y$ functions in the $I$ and $K$ Bessel functions, 
respectively. If we consider normal (nondivergent) initial conditions as a criterium
for physical solutions than this selects the expressions containing the $J$ and $I$
functions.

\noindent
\section{ Application to FRW barotropic cosmology}

Recently Faraoni \cite{Faraoni} showed that the equations
describing the FRW barotropic cosmologies can be combined in a 
simple Riccati equation of constant coefficients. 
In addition, Rosu \cite{Rosu} discussed in some detail the cosmological 
Riccati solutions and 
used nonrelativistic supersymmetry (generalized Darboux transformations)
to get cosmological Riccati equations of 
nonconstant coefficients.
Faraoni's Riccati equation is

\begin{equation}
\frac{dH}{d\eta }+cH^{2}=-kc,  \label{e39}
\end{equation}
where $H(\eta )=\frac{dR/d\eta }{R}$ is the Hubble parameter ($R$ is the scale
factor of the universe)
and $\eta$
is the conformal time, $c$ is related to the adiabatic index of the 
cosmological fluid under consideration, 
$c=\frac{3}{2}\gamma -1$. $k=0,-1,1$ are the curvature indices 
of the FRW universes, plane, open, and closed, respectively.

Applying the results of the previous section, for $a=c$, $b=-kc$, we get the
following solutions for the $\delta$- modified Hubble parameter:

\noindent For $k=1$ (the closed case), the $q$ parameter is real and we get 

\begin{equation} \label{h1}
H_{1}^{(+)}(\eta ; \delta )=\sqrt{\frac{1}{\Gamma (2-\delta ) }}\eta ^{%
\frac{1-\delta }{2}}\frac{J_{\frac{\delta -2}{3-\delta }}(\frac{2c}{3-\delta 
}\sqrt{\frac{1}{\Gamma (2-\delta )}}\eta ^{\frac{3-\delta }{2}%
})}{J_{\frac{1}{3-\delta }}(\frac{2c}{3-\delta }\sqrt{\frac{1}{\Gamma
(2-\delta )}}\eta ^{\frac{3-\delta }{2}})}~,
\end{equation}

\begin{equation}\label{h2}
H_{2}^{(+)}(\eta ; \delta )=\sqrt{\frac{1}{\Gamma (2-\delta )}}\eta ^{%
\frac{1-\delta }{2}}\frac{Y_{\frac{\delta -2}{3-\delta }}(\frac{2c}{3-\delta 
}\sqrt{\frac{1}{\Gamma (2-\delta )}}\eta ^{\frac{3-\delta }{2}%
})}{Y_{\frac{1}{3-\delta }}(\frac{2c}{3-\delta }\sqrt{\frac{1}{\Gamma
(2-\delta )}}\eta ^{\frac{3-\delta }{2}})}~.
\end{equation}

\noindent For $k=-1$ (the open case), the $q$ parameter is imaginary and we get

\begin{equation}\label{h3}
H_{1}^{(-)}(\eta ; \delta )=\sqrt{\frac{1}{\Gamma (2-\delta )}}
\eta ^{\frac{1-\delta }{2}}\frac{I_{\frac{\delta -2}{3-\delta }}
(\frac{2c}{3-\delta }\sqrt{\frac{1}{\Gamma (2-\delta )}}
\eta ^{\frac{3-\delta }{2}})}{I_{\frac{1}{3-\delta }}(\frac{2c}{3-\delta }
\sqrt{\frac{1}{\Gamma (2-\delta )}}\eta ^{\frac{3-\delta }{2}})},
\end{equation}

\begin{equation}\label{h4}
H_{2}^{(-)}(\eta ; \delta )=\sqrt{\frac{1}{\Gamma (2-\delta )}}
\eta ^{\frac{1-\delta }{2}}\frac{K_{\frac{\delta -2}{3-\delta }}
(\frac{2c}{3-\delta }\sqrt{\frac{1}{\Gamma (2-\delta ) }}
\eta ^{\frac{3-\delta }{2}})}{K_{\frac{1}{3-\delta }}(\frac{2c}{3-\delta }
\sqrt{\frac{1}{\Gamma (2-\delta )}}\eta ^{\frac{3-\delta }{2}})}.
\end{equation}

\noindent
The case $k=0$ corresponds to $b=0$, therefore it does not enter the present 
generalization in the sense that there is no change in the Riccati equation.

\begin{figure}[ht]
\centerline{\psfig{file=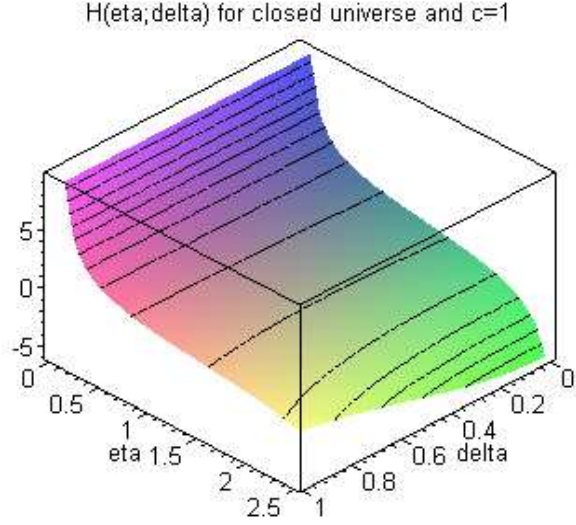,width=8.cm} }
\caption{Fractional Hubble parameter calculated according to the formula 
(41). }
\label{hclo1}
\end{figure}

\begin{figure}[ht]
\centerline{\psfig{file=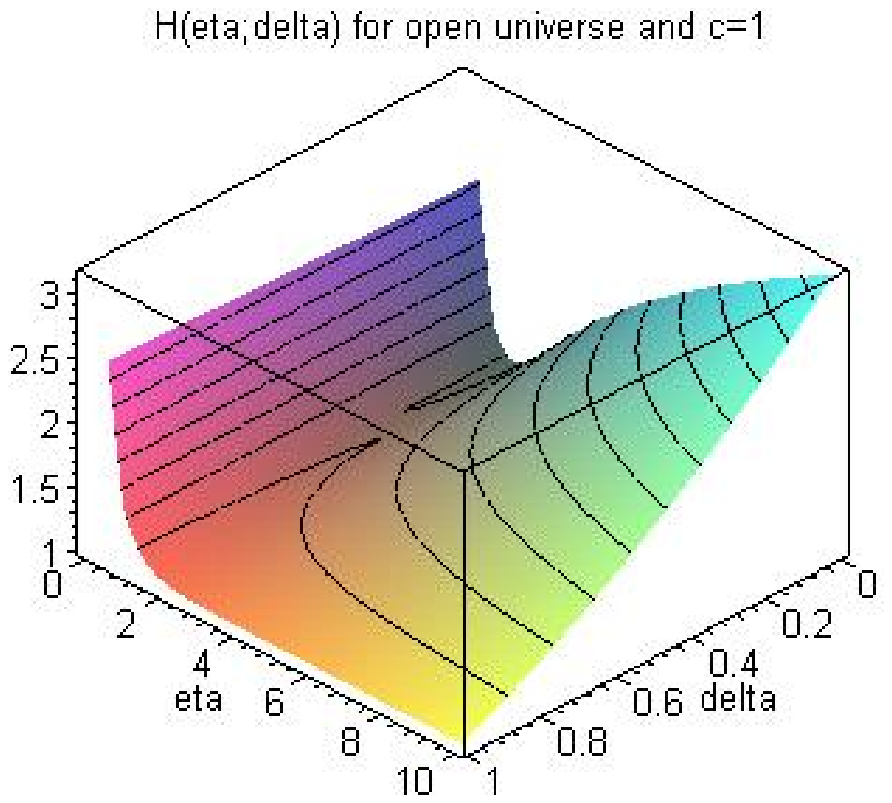,width=8.cm} }
\caption{Fractional Hubble parameter calculated according to the formula 
(43). }
\label{hopen}
\end{figure}

The formulas (\ref{h1}-\ref{h4}) can be considered a generalization of the results obtained by
 Faraoni. Nondivergent initial data correspond to (\ref{h1}) and (\ref{h3}). Three-dimensional plots
of these formulas are given in Figures (\ref{hclo1}) and (\ref{hopen}).
For $\delta =1$ we get $H_{1}^{(+)}(\eta ; \delta )=J_{-1/2}/J_{+1/2}={\rm cotan} (c\eta)$ and 
$H_{1}^{(-)}(\eta ; \delta )=I_{-1/2}/I_{+1/2}={\rm cotanh} (c\eta)$ that correspond
to the ordinary calculus. We mention that there are various works in the 
literature on the issue of geometric and physical interpretation of the 
fractional derivative and fractional integral, see, e.g., Podlubny \cite{P}.
In the cosmological case, the new parameter $\delta$ is introduced in the 
cosmological evolution of the Hubble parameter as a
consequence of applying a special fractional calculus to cosmological realms. 
In principle, as in statistical mechanics \cite{grig}, the fractional calculus can be considered 
as the  macroscopic manifestation of randomness. This has been argued to be so \cite{grig}
when there is no definite time-scale separation between the macroscopic and the microscopic 
level of description and this could be the case of cosmology itself. 

\end{document}